\documentclass[]{jfm}

\usepackage{graphicx}
\usepackage{newtxtext}
\usepackage{newtxmath}
\usepackage{natbib}
\usepackage{hyperref}
\hypersetup{
    colorlinks = true,
    urlcolor   = blue,
    citecolor  = black
}

\newcommand{\RomanNumeralCaps}[1]
\linenumbers

\usepackage{xcolor}

\title{
Coffee stain effect on a fibre from axisymmetric droplets}

\author{Marie Corpart\aff{1},
  Frédéric Restagno\aff{1}
 \and François Boulogne\aff{1}\corresp{\email{francois.boulogne@cnrs.fr}}}

\affiliation{\aff{1}Université Paris-Saclay, CNRS, Laboratoire de Physique des Solides, 91405, Orsay, France.}

\begin{document}
\maketitle

\begin{abstract}
The so-called coffee stain effect has been intensively studied over the past decades, but most of the studies are focused on sessile droplets. 
In this paper, we analyse the origin of the difference between the deposition of suspended particles in a sessile drop and in an axisymmetric drop deposited on a fibre.  
First, we model the shape of a drop on a fibre and its evaporative flux with some approximations to derive analytical calculations.
Then, for pinned contact lines, we solve the hydrodynamics equations in the liquid phase under the lubrication approximation to determine the flow velocity toward the contact lines. 
We comment these results by comparison to a sessile drop of similar evaporating conditions, and we show that the substrate curvature plays a role on the contact line depinning, the local evaporative flux, and the liquid flow field.
The competition between the advection and the Brownian motion indicates that the transport of the particles toward the contact line occurs in a volume localised in the close vicinity of the contact lines for a drop on a fibre.
Thus, the fibre geometry induces a weaker accumulation of particles at the contact line compared to a sessile drop, leading to the more homogeneous deposit observed experimentally.
\end{abstract}

\begin{keywords}
Authors should not enter keywords on the manuscript, as these must be chosen by the author during the online submission process and will then be added during the typesetting process (see \href{https://www.cambridge.org/core/journals/journal-of-fluid-mechanics/information/list-of-keywords}{Keyword PDF} for the full list).  Other classifications will be added at the same time.
\end{keywords}


%
%

\section{Introduction}

The deposition of a material on surfaces in a controlled manner is a key aspect in a broad range of applications. 
Among the coating techniques, the deposition of suspended particles, which can also be solute particles such as salts and polymers, through the evaporation of the carrying liquid here called the solvent is a common approach \citep{Routh2013,Brutin2018}.
When a volatile drop is deposited on a surface, particles are carried toward the contact line.
This so-called \textit{coffee stain effect} introduced by \cite{Deegan1997} and recently reviewed by \cite{Gelderblom2022} and \cite{Wilson2023}, is the consequence of a radial flow induced by evaporation. 
To rationalise this phenomenon, the fluid flow in the drop must be determined to explain the dynamics of the particle motion.
The resolution of the liquid flow is subordinated to the knowledge of the drop shape resulting from capillary phenomena, which is to a good approximation a spherical cap, for a drop on a flat substrate, when its size is smaller than the capillary length.
In addition, the derivation of the evaporative flux is necessary.
The evaporative flux of a circular disk, \textit{i.e.} a sessile drop with a vanishing contact angle, has been derived analytically by \cite{Cooke1967}, and generalisations to non-zero contact angle are also available in the literature \citep{Sreznevsky1882,Picknett1977}.
Thus, the fluid flow in the sessile drop has been derived theoretically through different contributions, \textit{e.g.} \cite{Deegan2000a,Hu2005,Popov2005,Zheng2009,Larson2014}.

In addition, \cite{Hamamoto2011} and then \cite{Marin2011, Marin2011a} revealed the rush-hour effect, which consists in an increase of the average particle velocity toward the contact line as the contact angle decreases in time. 
This particle velocity increase has an impact on the ordering of the particles in the final deposit~\citep{Marin2011, Marin2011a}.
The time evolution of the particle accumulation at the contact line has been satisfactorily predicted and measured by various authors \citep{Popov2005,Deegan2000,Monteux2011,Berteloot2012a,Larson2014,Boulogne2017a}.
The resulting deposition pattern forms a ring shape  \citep{Deegan2000,Routh2013,Brutin2018} that triggered various investigations to find strategies for tuning, limiting, or suppressing this effect.
For instance, studies focused on the liquid properties especially with solutal Marangoni flows \citep{Kajiya2009,Sempels2013,Kim2016,Pahlavan2021}, on the substrate hydrophobicity \citep{Gelderblom2011}, on the substrate permeability \citep{Boulogne2015b}, and on the multiple drops interaction \citep{Pradhan2015,Wray2020,Wray2021}.
See \cite{Mampallil2018} for a recent review.

Most of the literature is focused on drops on flat surfaces.
However, drops on fibres also represent a relevant situation for applications such as the drying of filters, clothing \citep{Duprat2022}, and insulating materials \citep{Sauret2015c}.
A liquid deposited on fibres can adopt a rich variety of morphologies.
We distinguish the clamshell shape, which corresponds to a small drop wetting a portion of the fibre perimeter from the barrel shape where the drop wets the fibre as a pearl on a necklace \citep{Chou2011}. 
More complex liquid shapes can be obtained on fibrous networks.
Fibres can be either crossed or parallel, which leads to a rich variety of equilibrium morphologies including liquid columns, distorted drops, and drops coexisting with columns \citep{Protiere2012,Sauret2015c}.

The deposition of particles dissolved in an evaporating clamshell drop on a fibre has already been investigated by \cite{Pham2002}, where a similar behaviour to sessile drops has been observed. 
In the barrel case, the liquid morphology is remarkable due to the substrate geometry. 
The fibre curvature induces an inflection point of the interface and a drop aspect ratio of order of the unity even for a perfectly wetting fluid in contrast to sessile drops \citep{Carroll1976,Brochard-Wyart1991,Lorenceau2006}.
By minimising the surface energy, \cite{Carroll1976} has obtained an analytical expression of the drop profile.
In addition, \cite{Corpart2022} recently obtained by numerical calculations the evaporative flux of an axisymmetric drop on a fibre, demonstrating that the divergence of the evaporative flux is localised near the contact line.
Thus, due to the differences in shape and evaporative flux induced by the substrate, we propose to investigate theoretically the particle deposition from a drop deposited on a fibre.
In this paper, our approach will favour analytical calculations when possible and we will limit ourselves to the regime where the contact line is pinned on the substrate.
Thus, we will rationalise the particle transport and the deposit left at the initial position of the contact lines, which will be compared to the well-established sessile drop.

To do so, we investigate theoretically the transport of particles due to evaporation in a drop wetting a fibre in an axisymmetric barrel configuration.
The obtained model is compared to the case of a sessile drop in terms of volume loss dynamics, characteristic velocities, and efficiency of the particles transport toward the contact line.
In Section \ref{sec:fibre}, we introduce the phenomenological equations for the drop shape and the evaporative flux in order to obtain analytical predictions.
Then, we derive the hydrodynamics equation in the liquid phase under the lubrication approximation to provide the velocity field toward the pinned contact line.
In Section \ref{sec:compare}, we compare the drop on a fibre with the sessile drop, and we comment about the main differences between these systems.
In Section \ref{sec:exp}, we present some qualitative experimental observations on the particle dynamics in both geometries that we compare to the theoretical investigations.

%
%
\section{Fluid flow of an evaporating drop on a fibre}\label{sec:fibre}

\begin{figure}
    \centering
    \includegraphics[width =0.9 \linewidth]{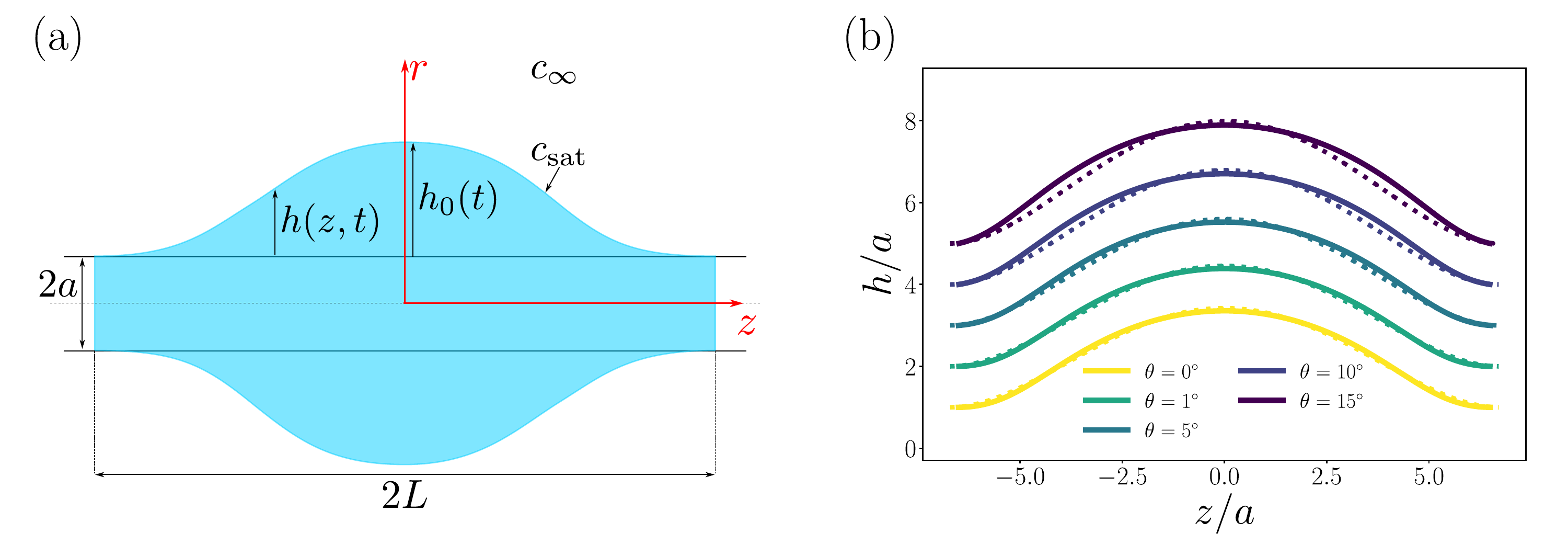}
    \caption{(a) Sketch presenting the notations of an axisymmetric drop of length $2L$ on a fibre of radius $a$. $c_{\rm sat}$ is the saturation vapour mass concentration, the vapour mass concentration in air at the liquid air/interface, and $c_\infty$ is the mass concentration of vapour far from the droplet.
    (b) Comparison of the analytical solution obtained by \cite{Carroll1976} of the dimensionless profile $h/a$ at $L/a = 6.5$ (solid lines), to the proposed ansatz~\eqref{eq:fibre-ansatz-profile-h_z_t} with $L/a \approx 6.7$ (dotted lines) for $\theta = 0,\, 1,\, 5,\, 10,\, 15^\circ$. 
    The apex height is obtained by fitting Carroll's profile by equation~\eqref{eq:fibre-ansatz-profile-h_z_t} and the length $L/a = 6.7$ is chosen to be the fitted length obtained for $\theta = 0$. 
    The dimensions chosen here corresponds approximately to a microlitre drop deposited on a glass fibre of a hundred micrometers radius. The curves are arbitrarily shifted for clarity.
    }
    \label{fig:notations_fibre_and_profiles_with_ansatz}
\end{figure}

We consider a volatile drop of density $\rho$, viscosity $\eta$, surface tension $\gamma$, and volume $\Omega$, on a fibre of radius $a$.
The contact angle of the liquid on the material, defined by Young's law, is denoted $\theta$.
On a horizontal single fibre, two configurations exist: a drop pierced by a fibre, the \textit{barrel} shape, or a drop wetting a portion of the fibre circumference, the \textit{clamshell} shape.
As shown by \cite{Chou2011}, this barrel shape is stable for a certain range of contact angles whose upper limit depends on the drop volume.
For sufficiently small contact angles, this configuration is therefore realistic.
Also, studies indicate that gravity, when sufficient, can off-centre the drop leading to an asymmetric shape \citep{Chou2011,Gupta2021}.
The role of gravity over capillarity can be quantified by the Worthington number ${\rm Wo}= \rho g \Omega / (\upi \gamma a)$ \citep{Worthington1885,Ferguson1912}.
We assume a negligible effect of the gravity, \textit{i.e.} ${\rm Wo} \ll 1$. 
The drop is therefore considered perfectly axisymmetric in this study. 

Thus, we adopt the cylindrical coordinate system shown in figure~\ref{fig:notations_fibre_and_profiles_with_ansatz}(a).
The profile of the liquid-vapour interface is described by a function $h(z,t)$.
The evaporation of the liquid generates a volume variation, and therefore an internal flow.
When the liquid is seeded with particles of radius $b$, the liquid flow can carry these particles.
Our aim is to predict analytically the particle transport in the liquid phase.
In the present work, we focus our interest on the regime for which the contact line is pinned, \textit{i.e.} the length $L$ is constant.
In practice, this pinning occurs due to the contact angle hysteresis and the additional pinning force due to the particles accumulated at the contact line creating defects \citep{Joanny1984,Meglio1992,Boulogne2016b}.
As evaporation proceeds, the drop height and the contact angle decrease.

The profile of the liquid interface \citep{Carroll1976} and evaporative flux \citep{Corpart2022} are particularly complex in this geometry.
Consequently, to perform an analytical analysis,  we will model the drop shape and the evaporative flux with phenomenological expressions, for which the validity will be discussed.
In addition, we will write the hydrodynamics equations to derive the main component of the liquid flow, along the fibre, while the contact lines remain pinned.
Then, we will be in position to discuss the particle transport in the drop.

\subsection{Profile of the liquid-vapour interface}

An analytical solution of the drop shape has been derived by \cite{Carroll1976}, but the elliptical integral involved in this solution would limit the analytical derivation of the present problem.
Thus, we choose to describe the drop profile with an ansatz fitting Carroll's solution

\begin{equation}\label{eq:fibre-ansatz-profile-h_z_t}
    h(z,t) = h_0(t) \left( 1 - \frac{z^2}{L^2}  \right)^2,
\end{equation}
where $h_0(t)$ is the drop height at the apex $z=0$ and $L$ is the distance from the apex to the contact line (Fig.~\ref{fig:notations_fibre_and_profiles_with_ansatz}(a).).
We consider the regime for which contact lines are pinned, so the length $L$ remains constant. 
The contact angle is defined by Young's law and depends only on the liquid/solid properties. 
For an axisymetric drop on a fibre, \cite{Carroll1976} demonstrated that the relation between the drop shape and the contact angle is not straightforward and in particular, the contact angle cannot be read directly from the slope between the interface and the surface of the fibre as it is for sessile drops. 
Typically, for contact angles between 0 and 30$^\circ$, the slope of the interface vanishes in the vicinity of the contact line as shown in figure~\ref{fig:notations_fibre_and_profiles_with_ansatz}(b) where we compare drop profiles obtained from \cite{Carroll1976} at $L/a = 6.5$ to equation \ref{eq:fibre-ansatz-profile-h_z_t} for $\theta = 0,\, 5,\, 10,\, \textrm{and}\, 15^\circ$. 
To describe the drop profile, we fit Carroll's profile with equation~\ref{eq:fibre-ansatz-profile-h_z_t} where $L$ and $h_0$ are the two adjustable parameters. 
To keep the wetted-length constant over the dynamics, we choose to set $L$ as the fitted length obtained for $\theta = 0$, $L/a \approx 6.7$.
Then, we use the values of $h_0$ obtained by the fit to calculate drop profile with equation~\eqref{eq:fibre-ansatz-profile-h_z_t} (see dashed lines in Fig.~\ref{fig:notations_fibre_and_profiles_with_ansatz}(b)). 
As shown in figure~\ref{fig:notations_fibre_and_profiles_with_ansatz}(b),
the proposed description of the drop profile is reasonably close to the analytical solution and describes its main features. 
We tested the validity of the ansatz for various drop profiles and we found a good agreement for drop with small contact angles, typically lower than 30$^\circ$ and dimensionless volume $\Omega/a^3 < 1000$.

The liquid volume is defined as
\begin{equation}\label{eq:fibre-def_volume}
    \Omega(t) = \int_a^{a+h(z,t)} \int_0^{2\upi}  \int_{-L}^L r \, {\rm d}r \,{\rm d}\theta \,{\rm d}z = \frac{32}{315} \upi L h_0(t) \left(21 a + 8 h_0(t)\right).
\end{equation}
With equation~\eqref{eq:fibre-ansatz-profile-h_z_t} and equation~\eqref{eq:fibre-def_volume}, we can estimate the volume of the drop and compare it to the volume calculated by \cite{Carroll1976} from his analytical description of the drop profile. 
The ansatz gives a good approximation of the drop volume with a relative error of about $10~\%$ for the largest contact angle tested, $\theta = 15^\circ$, for which the difference between the fit and the analytical profile is the largest.

\subsection{Evaporative flux}

The evaporation process is assumed to be limited by the diffusion of vapour in air \citep{Cazabat2010}, described by the vapour diffusion coefficient ${\cal D}_{\rm v}$.
The steady-state regime is reached on a timescale $L^2/{\cal D}_{\rm v}$, which is, in most situations, short compared to the observation timescale.
Thus, the vapour mass concentration field $c(r,z)$ is the solution of the Laplace equation $\triangle c = 0$.
The boundary conditions are $c=c_{\rm sat}$ at the liquid-vapour interface, $c=c_{\infty}$ far from the interface, and a no flux condition $\boldsymbol{n} \cdot \bnabla c = 0$ on the solid-gas interface characterised by its normal $\boldsymbol{n}$ (see Fig.~\ref{fig:notations_fibre_and_profiles_with_ansatz}(a)).
Once the mass concentration field is obtained, the evaporation velocity at the liquid-vapour interface is defined as $v_{\rm e}(z) = \frac{{\cal D}_{\rm v}}{\rho} \, \boldsymbol{n} \cdot \bnabla c|_{h(z,t)}$,  where the derivative is calculated at the liquid-vapour interface and $\rho$ is the density of the liquid. 

\begin{figure}
    \centering
    \includegraphics[width=0.7\linewidth]{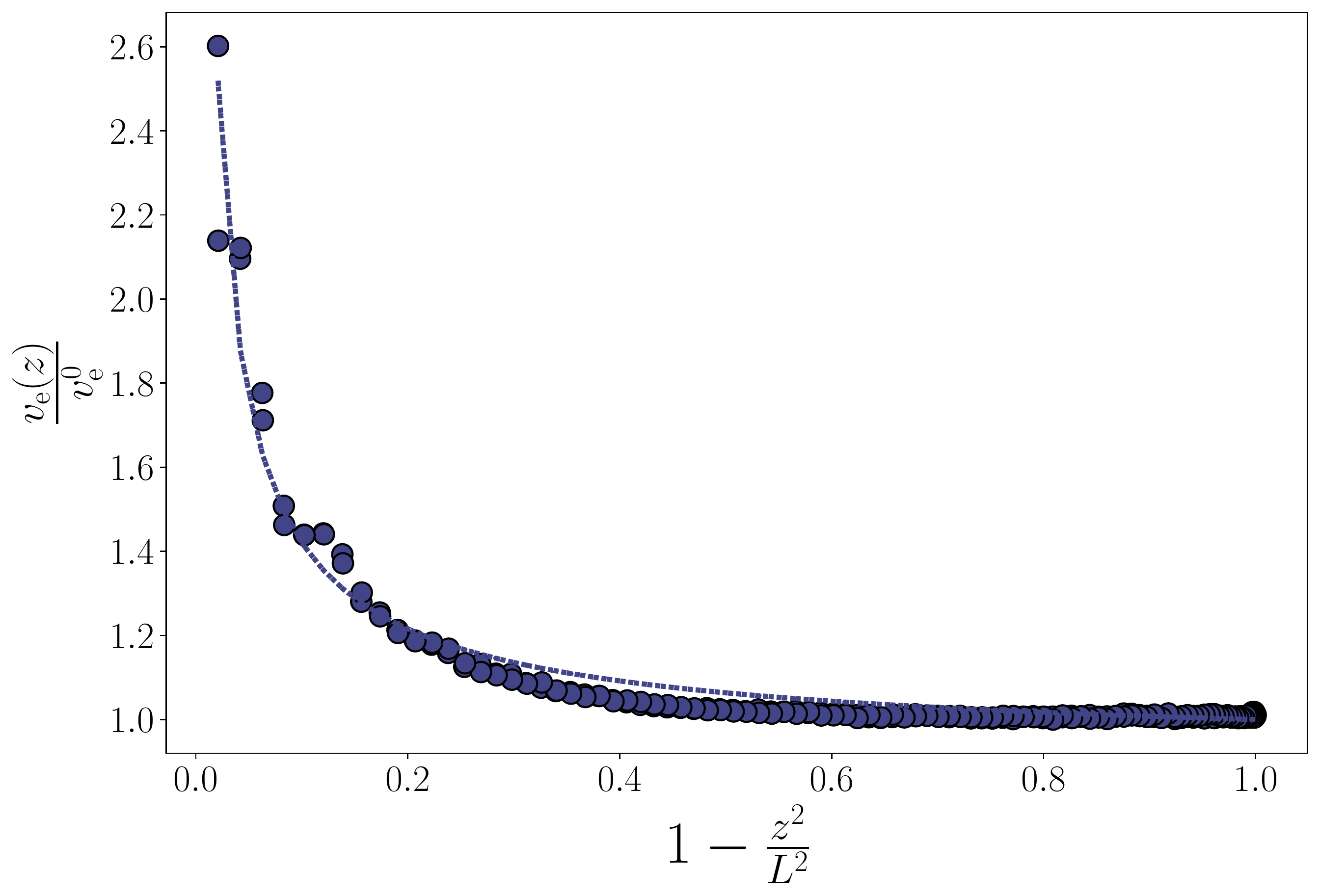}
    \caption{Local dimensionless evaporation velocity $v_{\rm e}(z)/v_{\rm e}^0$ as a function of $1-z^2 / L^2$ where $v_{\rm e}^0$ is the evaporation velocity at the drop apex.
    The points are obtained from numerical simulations using finite element methods presented in \cite{Corpart2022} for $a=125$~µm, $\theta = 10^\circ$, $\Omega = 1$~µL, which corresponds to $L/a \approx 6.7$.
    The dashed line is equation \eqref{eq:fibre-antsatz-evap-div} with the fitted parameters $\alpha= 0.7$ and $\beta= 0.1$.}
    \label{fig:evap_flux_simulations}
\end{figure}

\cite{Corpart2022} have recently performed numerical simulations using finite elements method to determine the local evaporative flux of a drop on a fibre.
In particular, this study revealed that the evaporation velocity differs from the well-known sessile drop in two aspects. 
First, the divergence is localised in the close vicinity of the contact line, and second, the contact angle has a weak effect on the evaporation velocity.
A typical result of these computations is presented in figure~\ref{fig:evap_flux_simulations}.
The localisation of the divergence near the contact line implies that the evaporation velocity cannot be written as a power law~\citep{Corpart2022}.
Instead, we propose to fit the results of the numerical simulations by
\begin{equation}\label{eq:fibre-antsatz-evap-div}
      v_{\rm e}(z)  = v_{\rm e}^0 \left[\beta\left( 1 - \frac{z^2}{L^2}  \right)^{-\alpha} + 1 -  \beta\right],
\end{equation}
where $\alpha$ and $\beta$ are constants, independent of the contact angle, for $\theta < 20^\circ$~\citep{Corpart2022}.
The prefactor $v_{\rm e}^0=v_{\rm e}(z=0)$ is the single parameter that depends on the environmental conditions and is proportional to ${\cal D}_{\rm v} (c_{\rm sat} - c_\infty) / \rho$.
The total evaporative flux writes $Q_{\rm e}(t) = \int v_{\rm e} {\rm d} S = 2\upi \int_{-L}^{L} v_{\rm e}(z) \, (a + h(z,t) ) {\rm d} z$, which leads after integration to
\begin{equation}\label{eq:fibre-Q_e_div-final}
Q_{\rm e}(t) =  2\upi v_{\rm e}^0 L \left(  A_1  a  + A_2  h_0(t)    \right),
\end{equation}
with 
\begin{subequations}
\begin{align}
A_1&=\sqrt{\upi}  \beta \frac{ \Gamma(1-\alpha) }{ \Gamma(\frac{3}{2} - \alpha)}   + 2 (1 - \beta),\\
A_2&=\sqrt{\upi}  \beta  \frac{ \Gamma(3-\alpha) }{ \Gamma(\frac{7}{2} - \alpha)}  + \frac{16}{15}  (1-\beta),
\end{align}
\end{subequations}
where $\Gamma$ is the Gamma-function \citep{Abramowitz1972}.

\subsection{Hydrodynamics}

\subsubsection{Lubrication approximation}

Now that the geometry and the evaporation dynamics  are described, we analyse the flow in the liquid phase.
By considerations of symmetries, the velocity field can be written in the cylindrical coordinates $(v_r(r, z), v_z(r, z))$.
The continuity equation is
\begin{equation}\label{eq:fibre-conv_continuity}
    \frac{1}{r}  \frac{\partial (r v_r)}{\partial r} + \frac{\partial v_z}{\partial z} = 0,
\end{equation}
and the Navier-Stokes equations in the stationary regime are
\begin{subequations}
\begin{align}
     -  \frac{\partial p}{\partial r} +
    \eta \left[\frac{1}{r}\frac{\partial}{\partial r}\left(r \frac{\partial v_r}{\partial r}\right)  + \frac{\partial^2 v_r}{\partial z^2}-\frac{v_r}{r^2} \right] = 0, \label{eq:fibre-NSfull_1}\\
    -  \frac{\partial p}{\partial z} + \eta \left[\frac{1}{r}\frac{\partial}{\partial r}\left(r \frac{\partial v_z}{\partial r}\right) + \frac{\partial^2 v_z}{\partial z^2}\right] = 0,\label{eq:fibre-NSfull_2}
\end{align}
\end{subequations}
where $p(r,z)$ is the liquid pressure.
The boundary conditions are no fluid slippage on the fibre and no stress at the liquid-vapour interface, \textit{i.e.}
\begin{subequations}
\begin{align}
    v_z(r,z) &= 0,\quad \mbox{on\ }\quad r=a \quad \mbox{and\ } \quad |z| \leq L, \label{eq:fibre-hydro_bc1}\\
   \frac{\partial v_z}{\partial r} &= 0,\quad \mbox{on\ }\quad r=a+h(z, t) \quad \mbox{and\ } \quad |z| \leq L.\label{eq:fibre-hydro_bc2}
\end{align}
\end{subequations}
Equation \ref{eq:fibre-hydro_bc2} means that tracers are assumed to have no surfactant effect, due to their size.
We apply the lubrication approximation to equations (\ref{eq:fibre-NSfull_1}, \ref{eq:fibre-NSfull_2}), which is valid for $\frac{h_0}{L} \frac{\rho h(z,t) v_z}{\eta} \ll 1$ \citep{Batchelor2000} and we get  

\begin{equation}\label{eq:fibre-NS-final}
    \frac{{\rm d} p}{{\rm d} z} = \eta \left[\frac{1}{r}\frac{\partial}{\partial r}\left(r \frac{\partial v_z}{\partial r}\right) \right].
\end{equation}

Based on the differential equation \eqref{eq:fibre-NS-final} and the boundary conditions \eqref{eq:fibre-hydro_bc1} and \eqref{eq:fibre-hydro_bc2}, we can calculate  the velocity field $v_z$

\begin{equation}\label{eq:fibre-vz_pressure}
    v_z(r, z, t) = - \frac{1}{\eta}  \frac{{\rm d} p}{{\rm d} z} \left[   \frac{ \left(a + h(z,t) \right)^2}{2} \ln\left(\frac{r}{r_0}\right) - \frac{r^2}{4}  \right],
\end{equation}
with $r_0 = a \exp\left( - a^2 / ( 2 (a+h(z,t))^2 ) \right)$.
We remark that the radial dependence of $v_z$ is not a quadratic function in contrast to the sessile drop that will be recalled thereafter in Section~\ref{sec:sessile}.

\subsubsection{Mass conservation}

To fully obtain the fluid velocity $v_z$, the pressure gradient ${\rm d} p /  {\rm d} z$ must be determined.
To do so, we write the mass conservation over a slice between $z$ and $z + {\rm d} z$,
\begin{equation}\label{eq:fibre-mass-conservation}
    \frac{\partial h}{\partial t} +  \frac{1}{2(a+h)}  \frac{\partial  }{\partial z} \left((2ah + h^2) \bar{v}_z  \right) + v_{\rm e}(z) =0,
\end{equation}
where $\bar{v}_z(z,t)$ is the velocity $v_z(r,z,t)$ averaged over a cross-section perpendicular to the fibre,
\begin{equation}\label{eq:fibre-mean_velocity-def}
    \bar{v}_z(z,t) =  \frac{2}{2ah+h^2} \int_a^{a+ h(z,t)} v_z(r,z,t)\, r \, {\rm d}r.
\end{equation}

Now, we can establish the relation between $\bar{v}_z$ and $v_z$.
First, we integrate equation \eqref{eq:fibre-mean_velocity-def} with equation \eqref{eq:fibre-vz_pressure} to express $\bar{v}_z$ as a function of ${\rm d} p /  {\rm d} z$.
Equation \eqref{eq:fibre-mean_velocity-def} is written
\begin{equation}\label{eq:fibre-mean_velocity_pressure}
    \bar{v}_z(z,t) = - \frac{1}{\eta}\frac{{\rm d} p}{{\rm d} z} \mathcal{G}(a,h(z, t)),
\end{equation}
where the geometrical function $\mathcal{G}(a,h(z, t))$ is given by

\begin{equation}
\mathcal{G}(a, h(z,t)) =\frac{ 1 }{4(2ah+h^2)} \left(   \frac{ a^4 }{2} +  (a+h)^4   \left[ \frac{a^2}{(a+h)^2} \left(  \frac{1}{2} - \ln \frac{a}{r_0} \right) + \ln \frac{a+h}{r_0} -1    \right]\right).
\end{equation}

Substituting the pressure derivative from  \eqref{eq:fibre-vz_pressure} in \eqref{eq:fibre-mean_velocity_pressure} yields 

\begin{equation}\label{eq:fibre-vz_final_form}
     v_z(r, z, t) =   \frac{ h \bar{v}_z}{\mathcal{G}(a, h(z,t))} \left[   \frac{ \left(a + h(z,t) \right)^2}{2} \ln\left(\frac{r}{r_0} \right) - \frac{r^2}{4}  \right].
\end{equation}

In the next subsections, we derive the time evolution of the drop profile $\partial h / \partial t$, which will be used along the evaporation velocity $v_{\rm e}(z)$ in the mass conservation \eqref{eq:fibre-mass-conservation} to obtain the average velocity $\bar{v}_z$ as a function of the evaporation dynamics.

\subsubsection{Liquid evaporation}

First, we calculate the time derivative of the liquid profile, $ \partial h / \partial t$, which appears in equation \eqref{eq:fibre-mass-conservation}.
The time derivative of the drop profile defined in equation \eqref{eq:fibre-ansatz-profile-h_z_t} considering a constant drop length $L$ gives
\begin{equation}\label{eq:fibre-height_rate}
    \frac{\partial h(z,t)}{\partial t} =   \frac{{\rm d} h_0 }{ {\rm d} t }    \left( 1 - \frac{z^2}{L^2}   \right)^2.
\end{equation}

In addition, the loss of liquid by evaporation compensates the total evaporative flux as ${\rm d} \Omega / {\rm d} t= - Q_{\rm e}(t)$.
Substituting the volume defined in equation \eqref{eq:fibre-def_volume}, we have

\begin{equation}\label{eq:fibre-apex_diff_eq}
     \frac{{\rm d} h_0}{{\rm d} t} = - \frac{315}{32 \upi}  \frac{Q_{\rm e}(t)}{ L (21 a + 16h_0(t)) },
\end{equation}
which fully defines the time derivative of the liquid profile.

Integrating equation~\ref{eq:fibre-apex_diff_eq} from 0 to $t$ and $h_{\rm i}$ to $h_0$ leads to 
\begin{equation}\label{eq:FIBRE_h_0_vs_t_implicite}
 \frac{16}{A_2} ( h_{\rm i} - h_0(t)) + \frac{a( 21 A_2 - 16 A_1)}{A_2^2} \ln{ \left( \frac{A_1 a + A_2 h_{\rm i}}{A_1 a + A_2 h_0(t)} \right)} = \frac{315}{16} v_{\rm e}^0 t.
\end{equation}
At the first leading order in $(h_{\rm i} - h_0)/h_{\rm i}$, we obtain the time variation of the height of the apex

\begin{equation}\label{eq:FIBRE_h_0_vs_t} 
     h_0(t) \approx h_{\rm i} - \frac{315}{16}\frac{A_1 a + A_2 h_{\rm i}}{21a + 16h_{\rm i}} v_{\rm e}^0 \, t. 
\end{equation}

\subsubsection{Mean liquid velocity}

Now, we derive the mean liquid velocity $\bar{v}_z$.
By substituting equations  \eqref{eq:fibre-height_rate} and \eqref{eq:fibre-apex_diff_eq}  in the mass conservation \eqref{eq:fibre-mass-conservation}, we have

\begin{equation}
   \frac{1}{2(a+h)} \frac{\partial   }{\partial z} \left( (2ah + h^2) \bar{v}_z \right)  =   \frac{315}{ 32 \upi   }  \frac{Q_{\rm e}(t)}{L(21a + 16 h_0)}   \left( 1 - \frac{z^2}{L^2}  \right)^2 - v_{\rm e}(z).
\end{equation}
The integration of this differential equation from $0$ to $z$ yields

\begin{equation}\label{eq:fibre-v_z_bar_Q_e}
    (2ah+h^2) \bar{v}_z = \, \frac{315}{16\upi}   \frac{Q_{\rm e}(t)}{  (21a + 16h_0)}  \frac{z}{L} \left[ a \, {\cal P}_1\left( \frac{z}{L} \right) + h_0\, {\cal P}_2\left( \frac{z}{L} \right) \right] -2 \int_0^z (a+h) v_{\rm e} {\rm d} z,
\end{equation}
where ${\cal P}_1(x) = 1 - \frac{2}{3} x^2  + \frac{1}{5}x^4$ and 
${\cal P}_2(x) = 1 - \frac{4}{3}x^2 + \frac{6}{5}x^4  - \frac{4}{7}x^6 + \frac{1}{9}x^8 $.

The plane perpendicular to the fibre at $z=0$ is a plane of symmetry, thus $ \bar{v}_z(z=0) = 0$.

From equation \eqref{eq:fibre-antsatz-evap-div}, we can calculate the remaining integral of equation \eqref{eq:fibre-v_z_bar_Q_e}

\begin{equation}
\begin{split}\label{eq:fibre_v_z_part_integral_divergence}
    2 \int_0^z (a+h) v_{\rm e}\, {\rm d}z &= 2 v_{\rm e}^0  z \left[ a \left(  \beta\, {}_2F_1\left(\frac{1}{2},\alpha;\frac{3}{2};\frac{z^2}{L^2}\right)  + (1-\beta) \right) \right. \\ &+ \left. h_0  \left(\beta\, {}_2F_1\left(\frac{1}{2},\alpha-2;\frac{3}{2};\frac{z^2}{L^2}\right)  + (1-\beta) \, {\cal P}_1\left( \frac{z}{L} \right)  \right) \right],
    \end{split}
\end{equation}
where the hypergeometric function ${}_2F_1(a,b;c;z)$ writes \citep{Abramowitz1972}
$$ {}_2F_1(a,b;c;z) = \sum_{n=0}^\infty \frac{(a)_n (b)_n}{(c)_n} \frac{z^n}{n!} = 1 + \frac{ab}{c}\frac{z}{1!} + \frac{a(a+1)b(b+1)}{c(c+1)}\frac{z^2}{2!} + \cdots.
$$

The combination of equations \eqref{eq:fibre-v_z_bar_Q_e} and \eqref{eq:fibre_v_z_part_integral_divergence} provides the mean velocity $\bar{v}_z(z,t)$.
The velocity profile $v_z(r,z,t)$ is therefore obtained by a straightforward substitution in equation \eqref{eq:fibre-vz_final_form}.
This description of the fluid flow permits to analyse the induced particle transport, which will be discussed in Section \ref{sec:compare}.

%
%
\section{Comparison between geometries}\label{sec:compare}

In this Section, we discuss the results obtained in Section \ref{sec:fibre} on a fibre that we compare to the well-established results for a drop on a flat surface (See Appendix \ref{sec:sessile}) \citep{Deegan2000,Popov2005,Stauber2014,Larson2014,Boulogne2017a}.
To adopt versatile notations for the two geometries, let $x$ be the $z$ or $r$ coordinate for the fibre or the sessile case, respectively.
Hence, the velocity toward the contact line is denoted $\bar{v}_x$.
Similarly, the length ${\cal L}$ denotes the length $L$ or $R$, respectively.

First, we compare the time evolution of the drop profiles for pinned contact lines to reveal the effect of the substrate curvature on the duration of this regime of evaporation with respect to the total time of evaporation.
Next, we analyse the fluid velocity toward the contact line and its efficiency to transport the particles through a Péclet number.
Finally, we compute to total number of particles accumulated at the contact line during the pinned regime.

\subsection{Methodology}\label{subsec:methodo}

To be able to compare the two geometries, a choice on the initial parameters has to be made. 
The possible parameters are  $\theta_{\rm i}$ the initial contact angle, $\cal L$ the wetted length, $\Omega(t=0)$ the initial volume, and $Q_{\rm e}(t=0)$, the initial evaporation rate. 
A common practice for comparison is to consider the same liquid-solid system such that the initial contact angle $\theta_{\rm i}$ is fixed. 
In addition, a similar evaporation rate $Q_{\rm e}$ for both systems brings the advantage of a comparable driving force for the particle transport. 
With equation~\eqref{eq:fibre-Q_e_div-final}, we can calculate the evaporation rate of a drop on a fibre whose geometry is given by equation~\eqref{eq:fibre-ansatz-profile-h_z_t} at $L/a \approx 6.7$ and $\theta = \theta_{\rm i} = 15^\circ$ (see Fig.~\ref{fig:notations_fibre_and_profiles_with_ansatz}(b)). 
Numerical simulation in a previous study by \cite{Corpart2022} provides $v_{\rm e}^0$  for a water droplet in the initial geometry described here and evaporating in dry air ($c_\infty = 0$) at 20~$^\circ$C ($c_{\rm sat} = 1.72 \times 10^{-2}$~kg/s and ${\cal D}_{\rm v} = 2.36 \times 10^{-5}$~m$^2$/s) \citep{Lide2008}.
We obtained $v_{\rm e}^0 \approx 4 \times 10^{-7}$~m/s. 
From equation~\eqref{eq:SESSILE_Q_e}, we can calculate the wetted length of the sessile droplet having the same evaporation rate for the same ambient conditions. 
We find a similar lengthscale $R \approx L$.
For instance, the barrel-shaped drop on a fibre of radius $a = 125$~µm has, initially, a wetted-length $L \approx 8.4 \times 10^{-4}$~m and a sessile drop evaporating at the same initial rate has a wetted length $R \approx 8.9 \times 10^{-4}$~m. 
We thus choose to make comparisons at the same wetted length $L/a \approx 6.7$ which is given by fitting Carroll's profile with equation~\eqref{eq:fibre-ansatz-profile-h_z_t} as explained in Section~\ref{sec:fibre} (see Fig.~\ref{fig:notations_fibre_and_profiles_with_ansatz}(b)). 
For the same wetted length and the same initial contact angle, the initial volume of the droplet is different in the two geometries. 
For example, for  $L = R \approx 8.4\times 10^{-4}$~m and $\theta_{\rm i} = 15^\circ$, the initial volume is $\Omega(t = 0) \approx 0.1$~µL for the sessile case and $\Omega(t = 0) \approx 0.6$~µL for the drop on fibre. 
The sessile drop has a smaller initial volume and thus a shorter lifetime than the drop on a fibre even if the initial evaporation rate is the same in the two geometries.
Additionally, since the particle transport depends on the particle size, we consider a particle diameter $2b=1$~µm motivated by the large number of studies on the coffee-stain using micrometer-sized particles.

To summarise, we choose to compare drops of same wetting length and same initial contact angle. 
This corresponds to drops having different initial volumes but the same initial evaporation rate for the two different geometries when evaporating in the same ambient conditions.   
In this framework, we can compare the mean velocities toward the contact line for the two systems.
The competition between the particle transport and the Brownian motion will be rationalised by a Péclet number.

\subsection{Time evolution of the drop shapes}

\begin{figure}
    \centering
    \includegraphics[width=\linewidth]{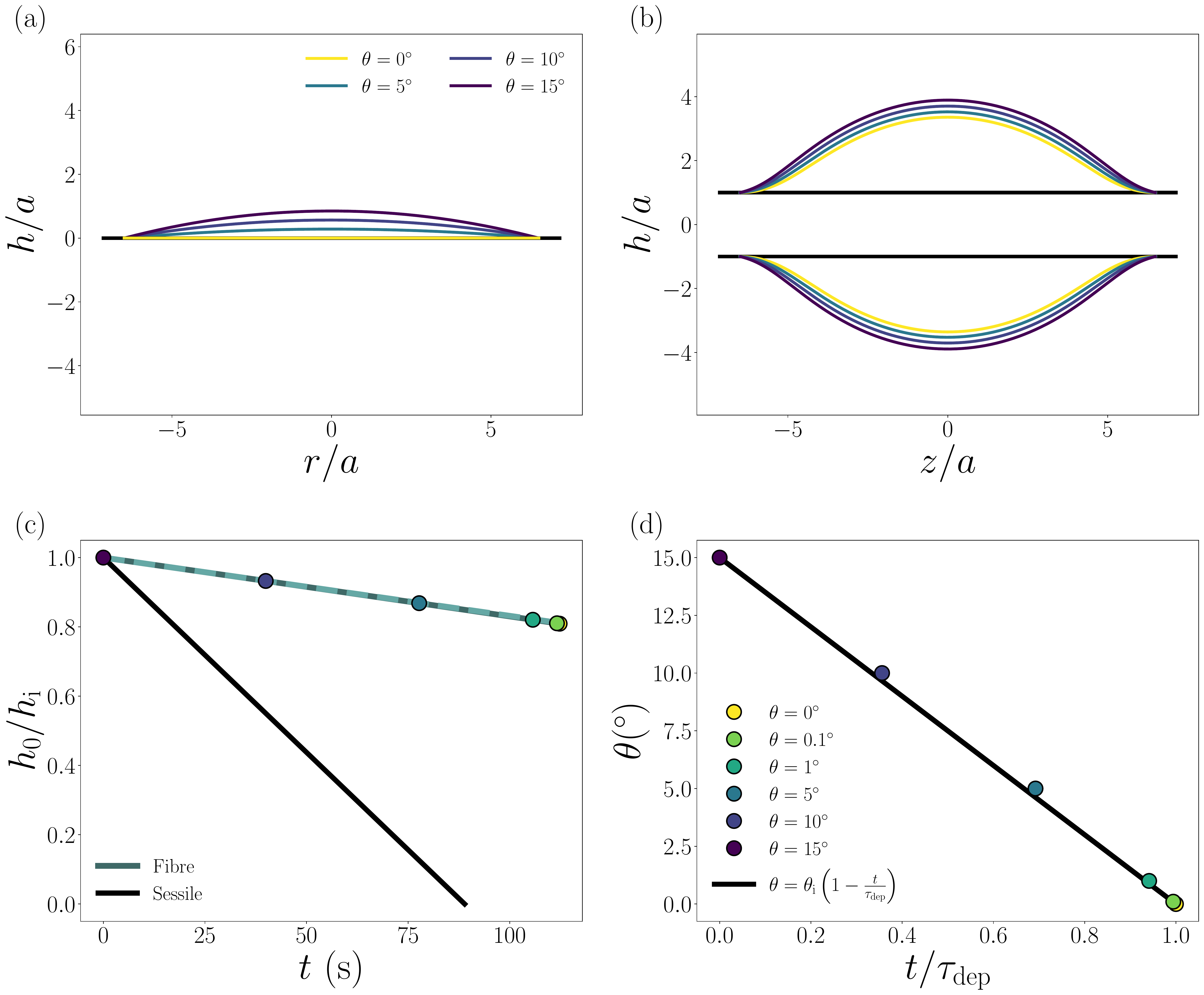}
    \caption{(a) -- (b) Evolution of the drop shape at different contact angles in (a) the sessile case (Eq.~ \ref{eq:SESSILE-profile}) and 
    (b) the fibre configuration for which the drop profile is described by elliptic integrals given by \cite{Carroll1976}. 
    (c) Time evolution of the drop height at the apex $h_0$ normalised by the initial height $h_{\rm i}$ for the two configurations. The black line corresponds to the sessile drop (Eq.~\ref{eq:SESSILE-apex}). 
    The blue lines are for an axisymmetric drop on a fibre, the solid line is equation~\ref{eq:FIBRE_h_0_vs_t_implicite} and the dashed line is the approximation given by equation~\ref{eq:FIBRE_h_0_vs_t}.
    (d) Dynamics of the contact angle $\theta$ for a sessile drop.
    (c) -- (d) Circles represent the apex heights (c) and contact angles (d) of the profiles of drops on fibres obtained with our ansatz (Eq.~\ref{eq:fibre-ansatz-profile-h_z_t}) and plotted in figure~\ref{fig:notations_fibre_and_profiles_with_ansatz}(b).
    Here, the comparisons are performed for the same wetted length ${\cal L}/a \approx 6.7$ and an initial contact angle $\theta_{\rm i} = 15^\circ$. This corresponds to water on glass substrate of initial volume $\Omega(t=0) \approx 0.6$~µL for the fibre configuration ($a = 125$~µm) and $\Omega(t=0) \approx 0.1$~µL for the sessile case.
    The initial evaporation rate $Q_{\rm e}(t=0)$ is nearly the same for the two configurations. For both geometries the ambient conditions are taken to be those of water evaporating in dry air at 20$^\circ$C, $c_{\rm sat} = 1.72 \times 10^{-2}$~kg/m$^{3}$, $ c_\infty = 0$ and ${\cal D}_{\rm v} = 2.36 \times 10^{-5}$~m$^2$/s~\citep{Lide2008}. 
    In this conditions, for a drop on a fibre ($\Omega = 1$~µL, $\theta = 10^\circ$ and $a = 125$~µm) we obtained by numerical simulations $v_{\rm e}^0 \approx 4 \times 10^{-7}$~m/s, $\alpha = 0.7$ and $\beta = 0.1$ (Fig.~\ref{fig:evap_flux_simulations}).
    }
    \label{fig:shape_temporal_variation}
\end{figure}

We analyse the evolution of the drop shape in both configurations.
The temporal evolution of drop heights, plotted in figure~\ref{fig:shape_temporal_variation}(c) are given by equation \ref{eq:FIBRE_h_0_vs_t_implicite} (blue solid line) approximated by equation~\ref{eq:FIBRE_h_0_vs_t} (dashed line) for the drop on a fibre and equation \ref{eq:SESSILE-apex} (black line) for the sessile drop.

The evolution of the drop shape for different contact angles for a sessile drop is illustrated in figure~\ref{fig:shape_temporal_variation}(a).
Due to the particle accumulation at the contact line that increases the pinning force \citep{Joanny1984,Meglio1992,Boulogne2016b}, we consider that the depinning nearly occurs  at a zero contact angle, which corresponds to a vanishing drop volume.
Now, considering the drop on a fibre, we plot in figure~\ref{fig:shape_temporal_variation}(b) the drop shape on a fibre having the same wetted length and the same contact angles as the sessile drops in figure~\ref{fig:shape_temporal_variation}(a). By fitting the profiles of figure~\ref{fig:shape_temporal_variation}(b)
with equation~\ref{eq:fibre-ansatz-profile-h_z_t} we get the apex heights and the corresponding times by inserting them in equation~\ref{eq:FIBRE_h_0_vs_t_implicite}. The results are represented by the circles in figure~\ref{fig:shape_temporal_variation}(c). This figure shows that the temporal evolution of the apex height of a drop on a fibre is well described by the approximated equation~\ref{eq:FIBRE_h_0_vs_t} (in dashed-line in Fig~\ref{fig:shape_temporal_variation}(c)) during the pinned-regime.

From figures~\ref{fig:shape_temporal_variation}(a) and (b), we also get the height of the apex at the depinning $h_0^{\rm dep} = h_0(\theta = 0)$. 
In the example studied here, we get $h_0^{\rm dep} = 0$ for the sessile drop and $h_0^{\rm dep} = 2.4a$ (corresponding to $h_0^{\rm dep} = 0.8 h_{\rm i}$) for the drop on a fibre. From that we get the duration of the pinned regime $\tau_{\rm dep}$ of the drop on a fibre by inserting $h_0^{\rm dep}$ into equation~\ref{eq:FIBRE_h_0_vs_t_implicite}.  In the example studied here, we find $\tau_{\rm dep} \approx 110$~s. 
In the case of the sessile drop the volume tends to zero at the end of the pinned regime, meaning that $\tau_{\rm dep} = \tau_{\rm e}$ (Eq.~\ref{eq:tau_evap_sessile}) the lifetime of a sessile drop evaporating entirely at constant contact radius. In the example studied here we find $\tau_{\rm e} \approx 90$~s. 
To establish the relationship between time and contact angle, we plot on figure~\ref{fig:shape_temporal_variation}(d) (circles) the temporal evolution of contact angle  obtained from the drop profiles represented in figure~\ref{fig:shape_temporal_variation}(b) for which we know $h_0$ and $\theta$. 
In solid black line, we represent $\theta(t) = \theta_{\rm i} \left(1 - {t}/{\tau_{\rm dep}}\right)$ valid for a sessile drop (Eq.~\ref{eq:SESSILE-apex}) which also describes very well the temporal evolution of the contact angle of a drop on a fibre. We thus find for both geometries that $h_0 \propto \theta \propto t$.

Moreover, as observed in figure~\ref{fig:shape_temporal_variation}(b, c), the relative variation of the drop height $h_0$ (and drop volume) during the pinned regime, is small for the drop on the fibre. 
During the pinned regime, a small amount of the initial volume has evaporated, which means that the duration of the pinned regime is short compared to the drop lifetime $\tau_{\rm dep} \ll \tau_{\rm e}$. 
The remarkable difference with the sessile drop is the significant liquid volume remaining at a zero contact angle, the contact angle at which the depinning is supposed to occur. 
We can estimate the remaining volume at the end of the pinned regime from equation \eqref{eq:fibre-def_volume}. 
For the geometry represented in figure~\ref{fig:shape_temporal_variation}(b), we find that only $35~\%$ of the initial volume has evaporated before the contact line depinning.

From this comparison, we can state that the geometry imposes constraints on the dynamics of the drop profile.
In particular, in contrast to the sessile drop, only a small fraction of the liquid volume can evaporate in the constant wetted length regime.
Also, on a fibre, a drop has two contact lines such that we expect that only one of the two lines depins.
At this time, the lower quantity of evaporated liquid implies that only a fraction of the particles are accumulated at the contact lines and that the complementary fraction of particles is still dispersed in the liquid phase. 
However to compute the number of accumulated particles at the contact line the velocity field inside the drop needs to be described which is the object of the next section.

\subsection{Transport of particles toward the contact line}
\subsubsection{Mean flow velocity toward the contact line}

\begin{figure}
    \centering
    \includegraphics[width=1\linewidth]{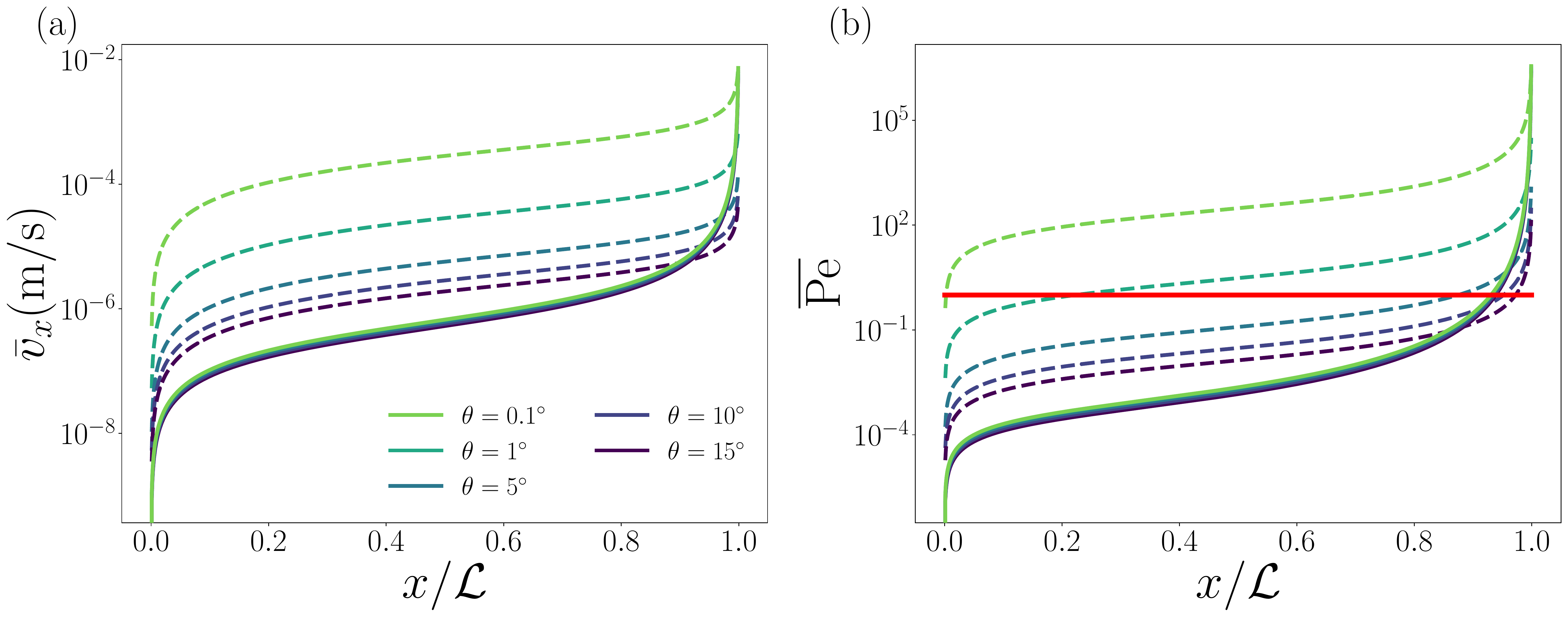}
    \caption{(a) Mean velocity of the flow toward the contact line $\bar{v}_x$ as a function of the dimensionless coordinate $x/{\cal L}$ along the solid surface. 
    The comparison is performed for water drops in both configurations with the same dimensionless wetted length $\mathcal{L}/a \approx 6.7$ and various contact angles (see caption). 
    Dashed lines correspond to $\bar{v}_r$ (Eq.~\eqref{eq:SESSILE_v_r_bar}), the average fluid velocity toward the contact line in sessile drops. 
    Solid lines are $\bar{v}_z$, the average fluid velocity in drop on fibre, given by equation~\eqref{eq:fibre-v_z_bar_Q_e}. 
    The dependence of $\bar{v}_z$ in $\theta$ is implicit and contained in the dimensions of the drop which are obtained by fitting Carroll's drop profile with equation~\eqref{eq:fibre-ansatz-profile-h_z_t} for each contact angle as done on Fig.~\ref{fig:notations_fibre_and_profiles_with_ansatz}(b). 
    (b) Péclet number $\overline{\rm Pe}$ deduced from the mean velocity and drop profile (Eqs.~\eqref{eq:fibre-ansatz-profile-h_z_t} and \eqref{eq:SESSILE-profile}) as a function of $x/{\cal L}$. 
    Solid lines correspond to drop on fibre and dashed lines to sessile drops having the same wetted length.
    The solid red line corresponds to $\overline{\rm Pe}=1$.
    For both geometries the ambient conditions are taken to be those of water evaporating in dry air at 20$^\circ$C, described in the paragraph~\ref{subsec:methodo} and in the figure~\ref{fig:shape_temporal_variation}.
    }
    \label{fig:velocity_toward_CL_and_Pe}
\end{figure}

In  figure~\ref{fig:velocity_toward_CL_and_Pe}(a), we plot $\bar{v}_x$ for both geometries, $x$ being either $z$ or $r$ according to the geometry. 
In the centre, $x=0$, the fluid velocity is equal to zero by symmetry.
We observe that, at the beginning of evaporation, \textit{i.e.} for $\theta = 15^\circ$ the flow towards the contact line is one order of magnitude higher in the sessile drop in most of the radial positions. 
Near the contact line, the mean liquid velocity diverges due to the vanishing liquid thickness for both geometries.
As the liquid evaporates \textit{i.e.} $\theta$ decreases, the so-called \textit{rush-hour effect}~\citep{Hamamoto2011,Marin2011} occurs in the sessile drop, \textit{i.e.} $\bar{v}_r$ increases in time.
However, $\bar{v}_z$  remains nearly constant for a drop on a fibre.
This difference is due to the curvature of the substrate that enables the existence of the peculiar axisymmetric barrel morphology, for which the variation of the drop profile remains limited when $\theta$ decreases (see Fig.~\ref{fig:shape_temporal_variation}(b)) unlike the case of the spherical cap on a flat substrate (Fig.~\ref{fig:shape_temporal_variation}(a)).

\subsubsection{Péclet number}

To describe the effective transport of particles toward the contact line, the action of the liquid flow on the particles must be compared to the Brownian motion. Particles are transported by the shear flow in the drop characterised by the mean shear rate  $\bar{\dot\gamma}  = \bar{v}(x,t) / h(x,t) $ and also by the Brownian motion ${\cal D} = k_{\rm B}T / (6\upi \eta b) $ where $k_{\rm B}$ is the Boltzmann constant and $T$ the temperature.
To compare these two competing forces, we introduce the mean Péclet number $\overline{\rm{Pe}}$  defined as $\overline{\rm Pe} = \bar{\dot\gamma} b^2 / {\cal D}$ \citep{Bossis1989}.

In figure~\ref{fig:velocity_toward_CL_and_Pe}(b), we plot the mean Péclet number $\overline{\rm Pe}$ as a function of the dimensionless position along the solid surface $x/{\cal L}$. 
As previously we observe that, initially, the mean Péclet number is comparable in the two geometries and diverges in the vicinity of the triple line where the liquid height vanishes. 
As the liquid evaporates the mean Péclet number strongly increases in the sessile case while it remains constant for the barrel-shaped drop on a fibre. 
Again, we attribute these differences to the difference of morphology between the droplets due to the curvature of the substrate.

\subsection{Number of particles accumulating at the contact line}

\subsubsection{Advection of a slice of fluid}

\begin{figure}
    \centering
    \includegraphics[width=.7\linewidth]{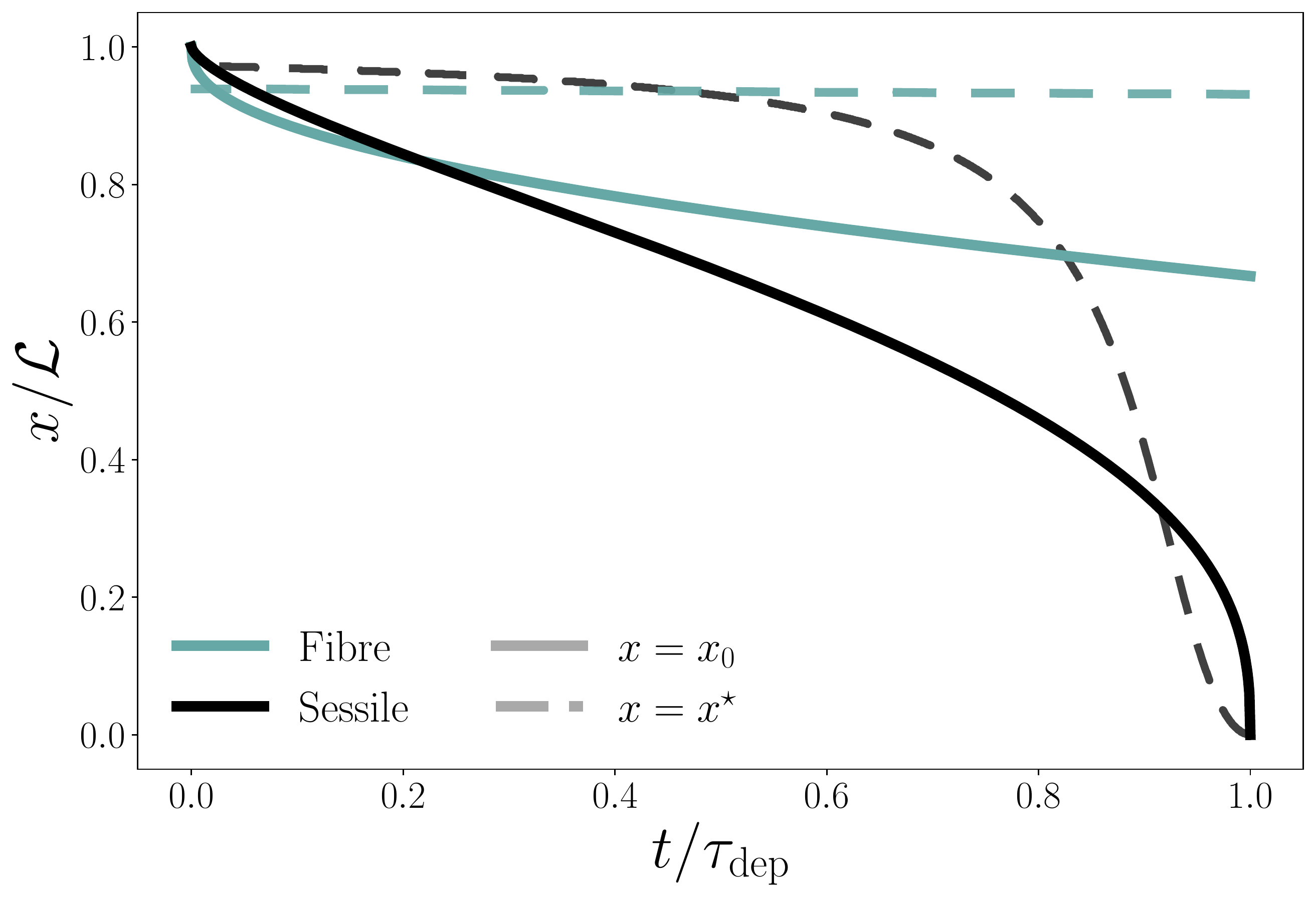}
    \caption{Temporal evolution of the dimensionless positions of the advected fluid layer $x_0/{\cal L}$ (solid lines) and of $x^\star/{\cal L}$ (dashed lines) the positions beyond which the P\'eclet number is greater than unity.
    Blue lines are used for axisymmetric drops on fibres, and black lines for sessile drops. Here, the comparison between the two geometries is performed in the conditions described in the paragraph~\ref{subsec:methodo} and in figure~\ref{fig:shape_temporal_variation}.
   }
    \label{fig:Peclet_domains}
\end{figure}

We consider a fluid layer at a position $x_0$ of an infinitesimal width ${\rm d}x$, which is advected toward the contact line at a velocity $\bar{v}_x(x_0, t)$.
The advection of a fluid layer is described by its position $x_0(t)$ that satisfies \citep{Deegan2000,Popov2005,Monteux2011,Berteloot2012a}

\begin{equation}\label{eq:FIBER_def_z_0}
    \frac{{\rm d} x_0}{{\rm d} t}  = \bar{v}_x(x_0, t).
\end{equation}

For the sessile drop, the solution is recalled in Appendix \ref{sec:sessile} (Eq.~\ref{eq:SESSILE_r0_solution}) and is represented in figure~\ref{fig:Peclet_domains} in solid black line.
For the fibre, the complexity of equation \ref{eq:fibre-v_z_bar_Q_e} makes us unable to find an analytical solution of the differential equation \eqref{eq:FIBER_def_z_0}.
Instead, we proceed to a numerical integration with \textsf{odeint} from scipy \citep{Jones2001}. The solution is plotted in solid blue line in figure~\ref{fig:Peclet_domains}.

We define $x^\star$ as the position along the solid surface for which $\overline{\rm Pe}(x^\star) = 1$. 
For $x > x^\star$, we have $\overline{\rm Pe} > 1$ such that the particles advection by the flow overcome their Brownian diffusion. 
The particles in the volume of liquid enclosed between $x = x^\star$ and $x = {\cal L}$ are advected by the flow and transported to the triple line.

In figure~\ref{fig:Peclet_domains}, we plot in dashed lines $x^\star/{\cal L}$ as a function of the dimensionless time for the two geometries. 
At the beginning of the drying, $x^\star/{\cal L}$ is closed to unity for both geometries. 
This means that the region over which the particles are transported by the liquid flow, corresponding to the region between $x =x^\star$ and $x= {\cal L}$, is localised in the close vicinity of the contact line at the beginning of evaporation. 
We even note that initially this region is smaller for a sessile drop than for a drop on a fibre $r^\star/R < z^\star/L$. Thus, at the beginning of evaporation, the transport of the particles toward the contact line is more efficient for the drop on a fibre. 
However, as the sessile drop evaporates, the region boundary position $r^\star/R$ continuously decreases to reach zero at the end of the pinned regime ($t = \tau_{\rm dep}$). 
On the fibre, $z^\star/L$ remains nearly constant, close to unity.
In other words, the width of the attraction zone in a sessile drop grows as the liquid evaporates.
Progressively, this zone occupies the entire drop, which leads to transport of the majority of the suspended particles toward the contact line. 
This is not the case for the drop on a fibre for which the zones in which the particles are effectively transported by the flow to the contact lines remain small and located in the close vicinity of the triple line.

The curves presented in figure \ref{fig:Peclet_domains} provide a comparison of the relative positions of $x^\star$ (dashed lines) and $x_0$ (solid lines).
During most of the pinned contact line regime, the area bounded by the position of the advected fluid layer $x_0$ includes small and large P\'eclet numbers domains ($x_0 < x^\star <\mathcal{L}$).
Exceptions are noticed at short timescales after evaporation starts and at the end of the pinned regime for the sessile drop.

\subsubsection{Particle accumulation dynamics in the large Péclet domain -- $x^\star < x_0$}\label{sec:accum_case1}

If $x^\star < x_0$, \textit{i.e.}  $\overline{\rm Pe} > 1$ between $x_0$ and $\mathcal{L}$, the particles in this layer are transported toward the contact line such that their number is conserved.
On the fibre, from the initial concentration (number of particles per unit volume) $c_{\rm i}$ and the initial liquid profile $h(z,\, t = 0)$, the number of particles $N_{{\rm CL}}$ accumulated at each contact line can be written
\begin{equation}\label{eq:FIBER-N_CL_integral_case1}
    N^{\rm fibre}_{{\rm CL}}(t) = c_{\rm i} \int_a^{a+h(z,\, t = 0)} \int_0^{2\pi} \int_{z_0(t)}^L r {\rm d} r \, {\rm d} \theta \, {\rm d} z,
\end{equation}
which gives,
\begin{equation}\label{eq:FIBER-N_CL_case1}
    N^{\rm fibre}_{\rm CL}(t) = \pi c_{\rm i}\, h_{\rm i}\, L\, \left(h_{\rm i}\, \mathcal{P}_3\left(\frac{z_0}{L}\right) + 2a\, \mathcal{P}_4\left(\frac{z_0}{L}\right) \right).
\end{equation}
with $\mathcal{P}_3(x) = - \frac{x^{9}}{9} + \frac{4 x^{7}}{7} - \frac{6 x^{5}}{5} + \frac{4 x^{3}}{3} - x + \frac{128}{315}$ and $\mathcal{P}_4(x) =- \frac{x^{5}}{5} + \frac{2 x^{3}}{3} - x + \frac{8}{15}$.
The above equation is shown as dashed blue line in figure~\ref{fig:COMPARAISON_N_CL_vs_t}.
The equivalent calculation for the sessile drop is recalled in Appendix \ref{sec:sessile-N_accum} (Eq.~\ref{eq:SESSILE-N_CL_case1}) and is represented in dashed grey line in figure~\ref{fig:COMPARAISON_N_CL_vs_t}.

\begin{figure}
    \centering
    \includegraphics[width = .7\linewidth]{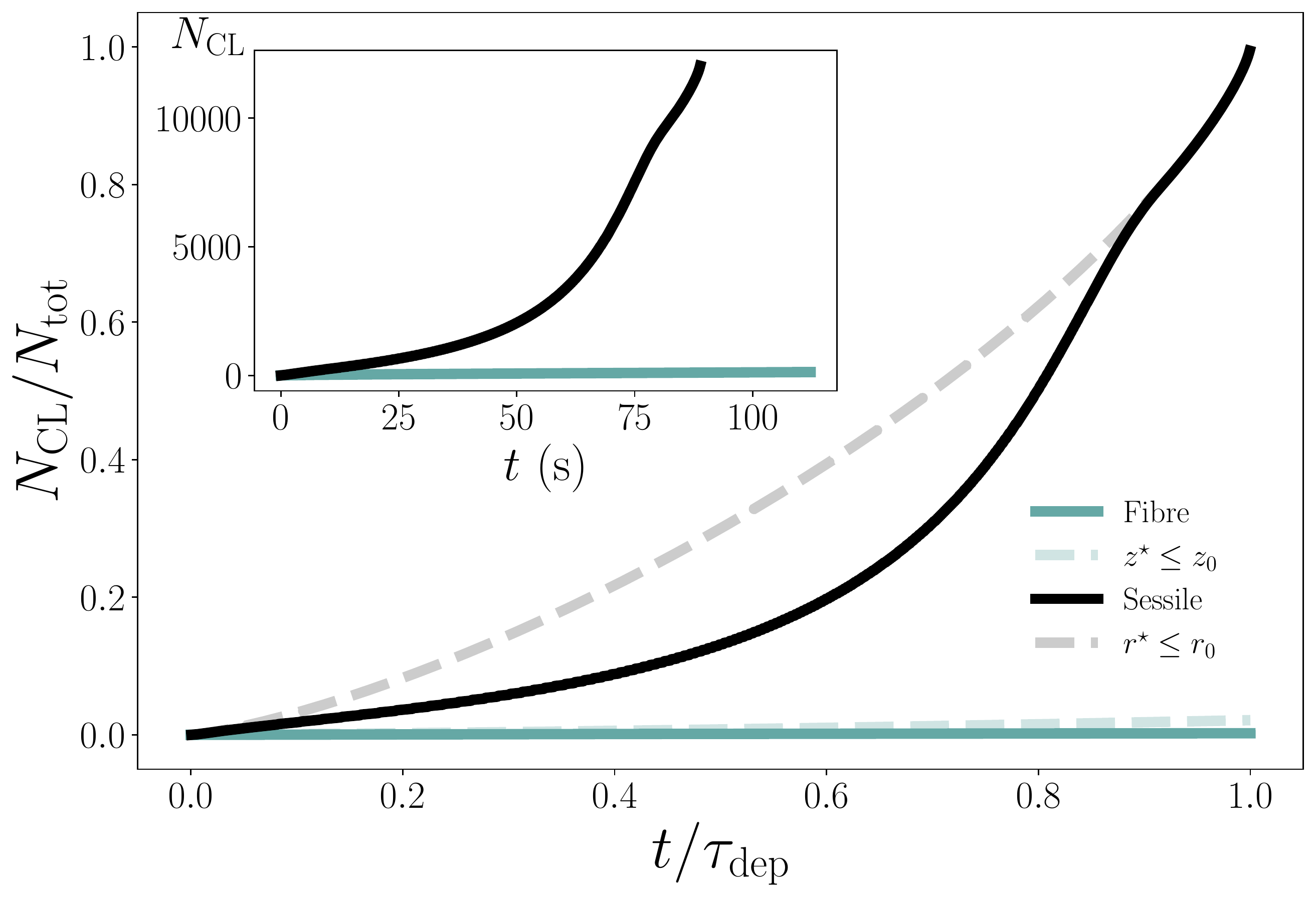}
    \caption{Time evolution of the dimensionless number of particles at the contact line $N_{\rm CL}/N_{\rm tot}$. Blue lines represent axisymmetric drops on fibres and black lines sessile drops.
    Solid lines are obtained from equation~\ref{eq:FIBER-N_CL_case1} (fibre) or equation~\ref{eq:SESSILE-N_CL_case1} (sessile) when $x_0 \geq x^\star$ and from equation~\ref{eq:FIBER-N_CL_case2} (fibre) or equation~\ref{eq:SESSILE-N_CL_case2} (sessile) when $x_0 \leq x^\star$.
    The dashed lines represent the results obtained under the assumption that all particles contained between $x_0$ and $\cal L$ are transported to the contact line and are plotted from Eq.~\eqref{eq:SESSILE-N_CL_case1} for a sessile drop (grey dashed line) and Eq.\eqref{eq:FIBER-N_CL_case1} for a drop on a fibre (blue dashed line). Here the comparison between the two geometries is performed in the conditions described in the paragraph~\ref{subsec:methodo} and in figure~\ref{fig:shape_temporal_variation}.
    The inset shows the temporal evolution of the number of particles $N_{\rm CL}$ plotted in solid lines in the main figure for an initial particle concentration $c_{\rm i} = 1 \times 10^8$ particles/mL.
    }
    \label{fig:COMPARAISON_N_CL_vs_t}
\end{figure}

\subsubsection{Particle accumulation dynamics over small and large Péclet domains -- $x^\star \geq x_0$}\label{sec:accum_case2}

Now, if positions $x^\star$ and $x_0$ are swapped, the number of accumulated particles is decomposed from two contributions.
The first contribution, from $x^\star$ to ${\cal L}$ is equivalent to equation \ref{eq:FIBER-N_CL_integral_case1}.
In this large Péclet domain, all the particles are advected with an increasing particle concentration as evaporation proceeds.
Between $x_0$ and $x^\star$, the small Péclet number indicates that Brownian motion maintains the particle concentration uniform.
Once the fluid layer reaches $x^\star$, particles are advected and concentrated as it is between $x^\star$ and ${\cal L}$.
Therefore, we evaluate the number of particles in the fluid layer at the position $x^\star$ with a particle concentration $c_{\rm i}$.
The sum of these two contributions gives

\begin{equation}
    N^{\rm fibre}_{{\rm CL}}(t) = c_{\rm i} \int_a^{a+h(z, t = 0)} \int_0^{2\pi} \int_{z^\star}^L r {\rm d} r \, {\rm d} \theta \, {\rm d} z + c_{\rm i} \int_a^{a+h(z^\star, t=0)} \int_0^{2\pi} \int_{z_0(t)}^{z^\star} r {\rm d} r \, {\rm d} \theta \, {\rm d} z.
\end{equation}
After integration, we obtain
\begin{equation}\label{eq:FIBER-N_CL_case2}
\begin{split}
    N^{\rm fibre}_{\rm CL}(t) = \pi c_{\rm i}\, h_{\rm i}\, L\, \left[h_{\rm i}\, \left(\mathcal{P}_3\left(\frac{z^\star}{L}\right) +  \left(1 - \left(\frac{z^\star}{L}\right)^2\right)^4 \, \left(\frac{z^\star}{L} - \frac{z_0}{L}\right) \right) \right. \\ \left. +  2a\, \left( \mathcal{P}_4\left(\frac{z^\star}{L}\right)  + \left(1 - \left(\frac{z^\star}{L}\right)^2\right)^2 \left(\frac{z^\star}{L} - \frac{z_0}{L}\right)  \right) \right].
\end{split}
\end{equation}
The same calculation is performed in Appendix~\ref{sec:sessile-N_accum} for the sessile drop.

Using the appropriate conditions according to the relative positions of $x_0$ and $x^\star$ shown in figure~\ref{fig:Peclet_domains}, we plot in solid lines in figure~\ref{fig:COMPARAISON_N_CL_vs_t} the dimensionless number of particles accumulated at the contact line. 
In the inset of figure~\ref{fig:COMPARAISON_N_CL_vs_t}, we plot the number of particles accumulated at the triple line over time for an initial  concentration of $c_{\rm i} =  1 \times 10^8$ particles/mL. 

Figure~\ref{fig:COMPARAISON_N_CL_vs_t} shows that the classical calculation  \citep{Deegan2000a,Berteloot2012a,Popov2005,Monteux2011,Boulogne2017a}, made in the literature for a sessile drop, valid if all the particles contained between $x_0$ and $\cal{L}$ are transported by the flow, leads to overestimate the number of particles accumulated at the triple line during the first part of the pinned regime. 
However, at the end of the drying the liquid height tends towards 0 which leads to $\overline{{\rm Pe}} > 1$ in almost the entire drop (cf. Fig.~\ref{fig:Peclet_domains}) \textit{i.e.} $r^\star \leq r_0$. 
The result is that almost all the particles are transported and deposited at the initial position of the contact line during drying, which corresponds to a typical density of $N_{\rm tot} / 2\pi R \approx 2$ particles/µm in the final deposit.
In practice, we must note also that the threshold value $\overline{{\rm Pe}} = 1$ is arbitrary and must be adjusted for a fine quantitative description.

For a drop on a fibre, on the other hand, $z^\star$ is almost constant and therefore during most of the pinned regime, $z^\star > z_0$ which means that the classical calculation of equation~\ref{eq:FIBER-N_CL_case1}, represented in blue dashed line in figure~\ref{fig:COMPARAISON_N_CL_vs_t}, overestimates the number of particles accumulated at the contact line. 
Indeed, taking into account the fact that Brownian diffusion dominates in the zone between $z_0$ and $z^\star$, we obtain a number of particles accumulated at the edge of the drop that is ten times lower than the one obtained by considering that all the particles between $z_0$ and $z^\star$ are advected by the flow.

Figure~\ref{fig:COMPARAISON_N_CL_vs_t} also shows that the particles contained in the drop on a fibre are transported toward the contact lines.
At the end of the pinned regime, the number of particles accumulated at the initial positions of the triple lines of a drop on a fibre corresponds to  $N_{\rm CL}(\tau_{\rm dep}) / 2 \pi a = 0.2$ particles/µm, which is about 10 times lower than the sessile drop. 
As shown in the inset, the duration of the pinned regime is approximately the same in both geometries $\tau_{\rm dep}^{\rm fibre} \approx 110$~s and $\tau_{\rm dep}^{\rm sessile} \approx 90$~s, but the rate of accumulation of particles at the contact line is lower in the drop on fibre than in the sessile drop.
This lower rate can be attributed to the overall lower fluid velocity and the narrow size of the advection-dominated domain in the fibre geometry.

Next we want to compare the results of the calculations with what is observed experimentally.

%
%
\section{Experimental observations}\label{sec:exp}

\begin{figure}
    \centering
    \includegraphics[width = 1\linewidth]{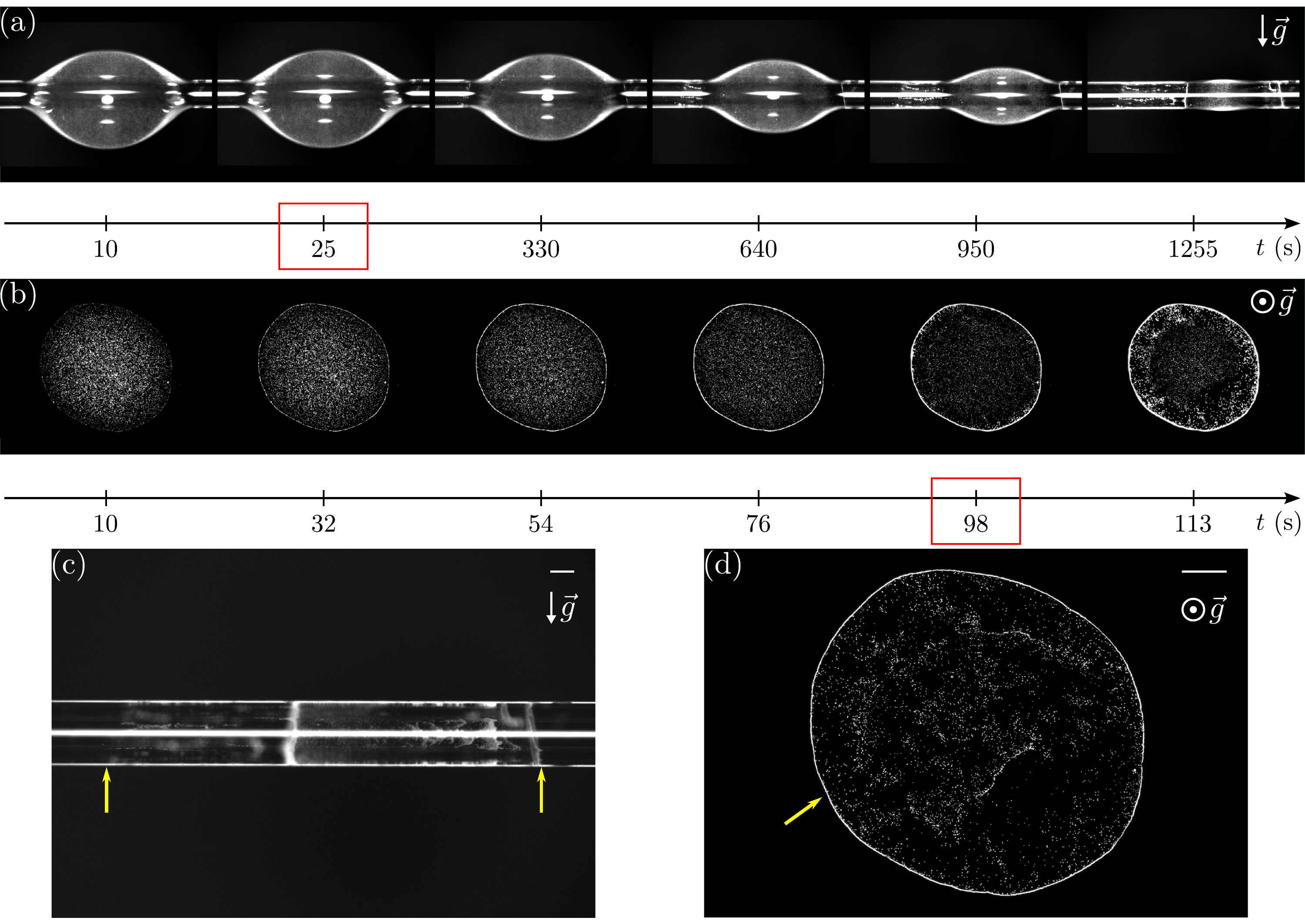}
    \caption{
    Experimental observations of particle transport induced by evaporation.
   (a,b) Temporal evolution of a drop on a fibre (side view) and on a flat surface (bottom view), respectively.
The initial wetted length is $R = L \approx 0.75$~mm. 
The times framed in red correspond to the moment when the depinning of one of the contact lines occurs.
(c,d) Photographs of the corresponding final deposit. 
The direction of gravity is shown in the pictures. The scale bars represent 0.2 mm and the yellow arrows indicate the initial position of the triple lines. 
Movies are provided in supplementary materials.
}\label{fig:experiment}
\end{figure}

\subsection{Materials and method}

We used fluorescent particles of polystyrene (Lifetechnologies) of diameter $2b = 1$ $\mu$m, diluted in pure water to a concentration of $c_{\rm i} = 1 \times 10^8$ particles/mL. 
The experiments are performed at $20~^\circ$C and at a relative humidity between
${\cal R}_{\rm H} = 30$~\% and ${\cal R}_{\rm H} = 47$~\%. 
The drops are deposited on the substrates using a micropipette (Eppendorf 0.1 - 2.5 $\mu$L). 
Images are recorded by using a camera (ORCA-Flash4.0, Hamamatsu).

For the sessile drop experiment, the drop is placed on a glass microscope slide washed with distilled water and soap and rinsed with acetone (Fisher, purity $\geq 99{,}8~\%$) and with anhydrous ethanol (Carlo Erba).
The observations are made from below using an inverted fluorescence microscope (IX83, Olympus) equipped with a 4$\times$ magnification objective (Olympus).
For the drop-on-fibre experiments, fibres of radius $a = 125$ $\mu$m are supplied by Saint-Gobain and activated by a plasma generator (Electro-Technics Products) prior to the experiments. 
The drop is observed from the side with a custom horizontal fluorescence microscope equipped with a 5$\times$ magnification objective (Mitutoyo).
The initial volumes are chosen to have the same initial wetted length in both geometries, such that 0.7~$\mu$L is deposited on the fibre and 0.1~$\mu$L on the microscope slide.

\subsection{Observations}

An example of the temporal evolution of the system is shown in figure \ref{fig:experiment}(a) for a drop on a fibre and \ref{fig:experiment}(b) for a sessile drop.
The Worthington number associated to the drop on the fibre is ${\rm Wo}\approx 0.2$, validating the small effect of gravity on the drop shape as observed in figure~\ref{fig:experiment}(a).

One of the main differences between the two geometries is that there is only one contact line in a sessile drop, whereas two independent contact lines exist in a drop on a fibre.
In both cases, the contact line depinning from its initial position is indicated by red frames in figures \ref{fig:experiment}(a) and \ref{fig:experiment}(b). 
The exact depinning time $\tau_{\rm dep}$ of one of the two triple lines of a drop on a fibre is difficult to determine experimentally due to the curvature of the substrate and the interface. 
However, it is observed that, on a fibre, the time during which the wetted length is constant is small compared to the total lifetime of the drop and represents about 1--10~\% of the lifetime. 
This is not observed for a sessile drop where the triple line remains pinned for the majority of the drying time, \textit{i.e.} 80--90~\% of the lifetime. 
The lifetime of the sessile drop is shorter than the lifetime of the drop on a fibre because its initial volume is lower. 
\cite{Corpart2022} have shown that the evaporative flux of a barrel-shaped droplet on a fibre is correctly approximated by the one of a spherical droplet in the same condition, such that the lifetime of a drop on a fibre can be estimated as $ \rho L^2 / (2 {\cal D}_{\rm v} [c_{\rm sat} - c_\infty])$. 
Comparing this result to the lifetime of a sessile drop defined in equation~\ref{eq:tau_evap_sessile} for the experimental condition tested here, we obtain $\tau_{\rm e}^{\rm fibre}/\tau_{\rm e}^{\rm sessile} \approx 8 /( \pi \theta_{\rm i}) \approx 10$ for $\theta_{\rm i} = 15^\circ$, which is in good agreement with what is measured experimentally as we get $\tau_{\rm e}^{\rm fibre}/\tau_{\rm e}^{\rm sessile} \approx 11$.

The final deposition is shown in figure \ref{fig:experiment}(c) and figure \ref{fig:experiment}(d) for these two geometries.
During the pinned regime of a drop on a fibre, it is observed that the areas over which particles are transported to the triple lines are small and remain localised in the vicinity of the contact line, so that few particles are deposited at the contact lines. 
Conversely, in a sessile drop evaporating at constant wetted length, we observe that the zone of attraction of the triple line progressively increases over time to finally invade the entire liquid. 
The particles are therefore mainly transported and deposited at the initial position of the contact line.

When one of the two contact lines unpins of the fibre, which is beyond the proposed model, the particles continue to accumulate at the stationary contact line, while the movement of the other triple line generates a complex flow in the drop. 
This contact line can then anchor again and a complex alternation of the movement of the left and right lines can be noticed, leading to the typical final morphology of the deposit observed in figure~\ref{fig:experiment}(c).

\subsection{Comparison with the proposed model}

There is a qualitative agreement between the observations and the predictions of the calculation. 
Indeed, for the drop on a fibre we observe experimentally that the zone over which the particles are transported towards the triple line is small and remains localised at the edge of the drops during the pinned regime and that there are few particles accumulated at the contact line during the pinned regime. 
The duration of the pinned regime is also short compared to the lifetime of the drop and when the depinning occurs, there is a significant volume of liquid remaining on the fibre.

However, we were unable to obtain a quantitative agreement between the theory and the experiments because of the difficulty of tracking the particles due to the curvature of the substrate and the liquid/air interface. 
In addition, the area where we can measure the velocity is located at the edge of the drop where the velocity diverges, so we cannot see any difference between the two geometries.

%
%
\section{Conclusions}

We conducted a theoretical investigation of the particle accumulation at the contact lines during the pinned contact line regime of an evaporating axisymmetric barrel-shaped drop on a fibre.
First, to obtain analytical expressions, we defined a phenomenological equation for the drop shape that we compared to the exact solution.
We used a phenomenological model for the evaporation velocity along the liquid-vapour interface, which is supported by a previous study \citep{Corpart2022}.
Within the lubrication approximation, we calculated the velocity field toward the contact line.
As the advection of particles competes with Brownian motion, we quantify the ability for the liquid flow to effectively transport the particles with a Péclet number. We compared our results to the well-known sessile drop geometry.

In our analysis, we highlighted that the liquid morphology is strongly different for both systems due to the fibre curvature.
A first consequence is that a large liquid volume remains on the fibre when one of the two contact lines unpins, whereas nearly all the sessile drop is evaporated.
A second consequence is that the divergence of the evaporation velocity is localised in the close vicinity of the triple line contrarily to the sessile drop.
From these two observations and our calculations, we have shown that the liquid flow velocity far from the contact line is order of magnitudes lower on the fibre, although the initial total evaporation rates are similar.
Nevertheless, the velocity field is not sufficient to obtain a description on the particle transport.
As the advection of particles  competes with Brownian motion, a Péclet number indicates how effective the particle transport is.
In a sessile drop, the domain where advection dominates diffusion grows in time, until a full invasion of the drop at the final stage, which is in perfect agreement with the common observation of an outward radial motion of particles.
On the fibre, the situation is strikingly different: advection remains located in a small region near the contact line for the same variation of the contact angle. Therefore the calculated rate of particle accumulation at the contact line is lower in a drop on a fibre than in a sessile drop.

As a result, the number of particles accumulated at the contact line at the end of the pinned regime on a fibre is weaker.
A unique and rich feature of the fibre geometry is the existence of two contact lines, topologically disconnected.
One of them remains pinned and particles keep accumulated in time, while the other one recedes, which brings an interesting dynamics, more complex than the sessile drop.
The evaporation dynamics continues with a succession of contact line pinning-depinning, leading to the succession of ring-like deposits shown in figure~\ref{fig:experiment}(c).
The fine description of the final pattern requires the modelling of the receding contact line \citep{Freed-Brown2014}.
Elucidating the receding dynamics of the contact lines induced by evaporation coupled with the particle deposition is an
important consideration for future studies.

\backsection[Supplementary data]{\label{SupMat}Supplementary material and movies are available at }

\backsection[Acknowledgements]{We thank Saint-Gobain and ANRT for funding this study and J. Delavoipière and M. Lamblet for useful discussions.}

\backsection[Declaration of interests]{The authors report no conflict of interest.}

\appendix

%
%
\section{Fluid flow of an evaporating sessile drop}\label{sec:sessile}

In this Appendix, we recall some results formerly obtained on the evaporation of a sessile drop with a pinned contact line.
These results will be used for quantitatively comparing the evaporation of sessile drop and a drop on a fibre in the next Section.

We consider a drop of volatile liquid of the same properties as in the main text and we recall some results in the former works of \cite{Deegan2000,Popov2005,Stauber2014,Larson2014,Boulogne2017a} and summarised recently by \cite{Gelderblom2022}. 
The drop is sitting on a flat surface with a contact angle $\theta$ and a constant contact radius $R$. We consider small volume such that the gravitational effects are negligible and the geometry of the liquid is well described by a spherical cap.
We assume a small contact angle, which simplifies the description of the drop shape and the evaporative flux.
Thus, the drop profile is 
\begin{equation}\label{eq:SESSILE-profile}
        h(r,t) = h_0(t) \left( 1 - \frac{r^2}{R^2} \right),
    \end{equation}
where $h_0(t) \approx R \theta(t)/2 $ is the height at the apex. 
In the pinned regime,

\begin{equation}\label{eq:SESSILE-apex}
        h_0(t) =h_{\rm i} \left(1 - \frac{t}{\tau_{\rm dep}}\right),
\end{equation}
with $h_{\rm i}$ the initial height of the apex and $\tau_{\rm dep}$ the duration of the evaporation process in the pinned regime.
For a sessile drop, the receding contact angle tends to zero, in particular due to the presence of particles \citep{Joanny1984,Meglio1992,Boulogne2016b}.
Therefore, we can estimate with a good approximation that $\tau_{\rm dep} \approx \tau_{\rm e}$, the lifetime of the drop. 

In these conditions, the evaporation velocity is

\begin{equation}
    v_{\rm e}(r) = v_{\rm e}^0  \left( 1 - \frac{r^2}{R^2} \right)^{-1/2},
\end{equation}
with the characteristic evaporation velocity $v_{\rm e}^0 = 2{\cal D}_{\rm v} (c_{\rm sat} - c_\infty)/\left(\upi \rho R\right)$, where ${\cal D}_{\rm v}$ is the diffusion coefficient of the vapour in the air. 
Then, the total evaporative flux is
\begin{equation}\label{eq:SESSILE_Q_e}
    Q_{\rm e} = 4 \mathcal{D}_{\rm v} (c_\mathrm{sat} - c_\infty) R.
\end{equation}
From equation~\eqref{eq:SESSILE_Q_e}, we can write the evaporative time assuming that all the liquid evaporates during the pinned-regime
\begin{equation}\label{eq:tau_evap_sessile}
    \tau_{\rm e} = \frac{\upi \rho h_{\rm i} R}{8 {\cal D}_{\rm v} (c_{\rm sat} - c_\infty)}.
\end{equation}
Under the lubrication approximation, the mean radial velocity is

\begin{equation}\label{eq:SESSILE_v_r_bar}
    \bar{v}_r(r,t) = v_{\rm e}^0 \frac{R^2  }{rh(r,t)}  \left( \left(1 - \frac{r^2}{R^2}\right)^{1/2} - \left(1 - \frac{r^2}{R^2}\right)^2\right),
\end{equation}
and the radial velocity field corresponds to half of a Poiseuille flow, given by

\begin{equation}\label{eq:SESSILE_v_r}
    v_r(r,z,t) =  \frac{3}{2}  \frac{R^2 v_{\rm e}^0}{r h(r,t)^3}  \left( \left( 1 - \frac{r^2}{R^2} \right)^{1/2} - \left(1 - \frac{r^2}{R^2}\right)^2\right) \left(  z^2 - 2h(r,t)z \right).\
\end{equation}

\subsection{Particle accumulation dynamics}\label{sec:sessile-N_accum}

The number of particles $N_{\rm{CL}}(t)$ accumulating at the contact line is the sum of the particles contained in the volume between $r_0(t)$ and $R$ where $r_0(t)$ is defined as
        
        \begin{equation}\label{eq:SESSILE_def_r_0}
            \frac{{\rm d} r_0}{{\rm d} t}  = \bar{v}_r(r_0(t),t),
        \end{equation}
which leads after integration to
          \begin{equation}\label{eq:SESSILE_r0_solution}
            \frac{r_0(t)}{R} = \sqrt{ 1 - \left(1-\left(1-\frac{t}{\tau_e}\right)^{3/4}\right)^{2/3}}.
        \end{equation}

As presented in Section~\ref{sec:accum_case1}, the number of accumulated particles in the case $r^\star < r_0$ writes
        
        \begin{equation}\label{eq:SESSILE_number_particles_integral_case1}
            N^{\rm sessile}_{{\rm CL}}(t) = 2\pi c_{\rm i} \int_{r_0(t)}^R   h(r', t=0)\, r' \mbox{d}r',
        \end{equation}
        where $c_{\rm i}$ is the initial particle concentration.

        \begin{equation}\label{eq:SESSILE-N_CL_case1}
            N^{\rm sessile}_{{\rm CL}}(t) = 2\pi c_{\rm i} h_{\rm i} R^2 \left(\frac{1}{4} - \frac{1}{2} \frac{r_0(t)^2}{R^2} + \frac{1}{4}  \frac{r
_0(t)^4}{R^4}\right),
        \end{equation}

For the other case where $r^\star > r_0$, we have

\begin{equation}
    N^{\rm sessile}_{{\rm CL}}(t) = c_{\rm i} \int_{ r^\star}^{R} \int_0^{ 2\pi} \int_{0}^{h(r, t=0)} r {\rm d} r \, {\rm d} \theta \, {\rm d} z + c_{\rm i} \int_{r_0}^{r^\star} \int_0^{2\pi} \int_{0}^{h(r^\star, t = 0)} r {\rm d} r \, {\rm d} \theta \, {\rm d} z,
\end{equation}
which writes after integration:

\begin{equation}\label{eq:SESSILE-N_CL_case2}
    N^{\rm sessile}_{{\rm CL}}(t) = 2 \pi c_{\rm i} h_{\rm i} R^2\left[ \frac{1}{4}  - \frac{1}{4}  
    \frac{r^\star(t)^4}{R^4} - \frac{1}{2}\frac{r_0^2(t)}{R^2} \left(1 - \frac{r^\star(t)^2}{R^2} \right) \right].
\end{equation}

\bibliography{biblio}

\bibliographystyle{jfm}

\end{document}